\documentclass[superscriptaddress,reprint,secnumarabic,amssymb]{revtex4-1}

\usepackage{gensymb}

\usepackage{graphics}
\usepackage{epsfig}
\usepackage{epstopdf}
\usepackage{amsmath}
\usepackage{subfigure}
\usepackage{float}
\usepackage{enumerate}
\usepackage{array}

\begin{document}

\title{Axial Super-resolution Evanescent Wave Tomography}

\author{Sarang Pendharker}
\affiliation{Department of Electrical and Computer Engineering, University of Alberta, Edmonton, Alberta T6G 1H9, Canada}
\email{sarang.pendharker@ualberta.ca}

\author{Swapnali Shende}
\affiliation{Department of Chemical and Materials Engineering, University of Alberta, Edmonton, Alberta T6G 1H9, Canada}

\author{Ward Newman}
\affiliation{Department of Electrical and Computer Engineering, University of Alberta, Edmonton, Alberta T6G 1H9, Canada}

\affiliation{Birck Nanotechnology Center, School of Electrical and Computer Engineering, Purdue University, West Lafayette, IN 47906, USA }

\author{Stephen Ogg}

\affiliation{Department of Medical Microbiology $\&$ Immunology, University of Alberta, Edmonton, Alberta T6G 2E1, Canada}

\author{Neda Nazemifard}

\affiliation{Department of Chemical and Materials Engineering, University of Alberta, Edmonton, Alberta T6G 1H9, Canada}

\author{Zubin Jacob}
\affiliation{Department of Electrical and Computer Engineering, University of Alberta, Edmonton, Alberta T6G 1H9, Canada}
\affiliation{Birck Nanotechnology Center, School of Electrical and Computer Engineering, Purdue University, West Lafayette, IN 47906, USA }



\begin{abstract}
	Optical tomographic reconstruction of a 3D nanoscale specimen is hindered by the axial diffraction limit, which is 2-3 times worse than the focal plane resolution. We propose and experimentally demonstrate an axial super-resolution evanescent wave tomography (AxSET) method that enables the use of regular evanescent wave microscopes like Total Internal Reflection Fluorescence Microscope (TIRF) beyond surface imaging, and achieve tomographic reconstruction with axial super-resolution. Our proposed method based on Fourier reconstruction achieves axial super-resolution by extracting information from multiple sets of three-dimensional fluorescence images when the sample is illuminated by an evanescent wave. We propose a procedure to extract super-resolution features from the incremental penetration of an evanescent wave and support our theory by 1D (along the optical axis) and 3D simulations. We validate our claims by experimentally demonstrating tomographic reconstruction of microtubules in HeLa cells with an axial resolution of $\sim$130 nm. Our method does not require any additional optical components or sample preparation. The proposed method can be combined with focal plane super-resolution techniques like STORM and can also be adapted for THz and microwave near-field tomography. 
\end{abstract}

\maketitle

Optical tomography is a major tool in three-dimensional visualization of sub-micrometer scale specimens in biology, material sciences and nano-fabrication technology \cite{osseforth2014simultaneous,huang2008three,xu2012dual}. Tomographic reconstruction is done by optical sectioning of the object in the focal plane followed by 3D stitching of the acquired z-stack of focal plane images. Resolution of the 3D tomographic reconstruction of an object is therefore governed by the focal plane resolution and the axial resolution of the underlying optical image acquisition. A wide range of well advanced fluorescence based super-resolution microscopy techniques have been reported and are currently in practice \cite{huang2010breaking}. Super-resolution in the focal plane can be achieved by localization techniques like Stimulated Emission Depletion (STED) \cite{hell1994breaking}, Stochastic Optical Reconstruction Microscopy (STORM) \cite{thompson2002precise}, Photo-Activated Localization Microscopy (PALM) \cite{patterson2010superresolution}, or by patterned illumination techniques like Structured Illumination Microscopy (SIM) \cite{gustafsson2000surpassing}, Plasmonic Structured Illumination Microscopy (PSIM) \cite{wei2014wide} etc. Even though many focal-plane super-resolution techniques are available, accurate 3D tomographic reconstruction still remains a challenge due to the ellipsoidal shape of the Point Spread Function (PSF), which makes resolution in axial direction 2-3 times worse \cite{hell2007far}.

The 4PI \cite{hell1992properties} and I$^5$M microscope \cite{gustafsson1999i5m} have increased the resolution in the axial direction with an almost spherical PSF. More recently a triple-view capture and fusion approach \cite{wu2016simultaneous} has been reported to improve volumetric resolution by a factor of two. However, these techniques require imaging and illuminating the same focal plane from both sides of the sample and depend on extensive optical components and precise optical phase matching. 3D STED \cite{osseforth2014simultaneous} and 3D STORM \cite{huang2008three,xu2012dual} for three-dimensional localization have also been reported. Axial super-resolution can also be achieved by placing a reflective mirror behind the sample to squeeze the PSF in the axial direction by interference from the reflected STED beam \cite{deguchi2014axial,yang2016mirror}. However, STORM and STED impose constraints on properties of the fluorescent probes, limiting their applicability to image photo-switchable fluorophores and samples with a sharp emission spectrum, respectively. Evanescent wave illumination techniques like Total internal Reflection Fluorescence (TIRF) microscopy \cite{axelrod2008total,shen2014tirf}, Plasmon enhanced TIRF \cite{balaa2009surface}, variable-angle TIRF \cite{wan2011variable} and pseudo-TIRF \cite{sun2011whole} provide super-resolution along the optical axis with very high signal-to-noise ratio (SNR) without imposing constraints on the fluorescence properties of the sample. However, the capabilities of the evanescent wave techniques were so far limited to near-surface imaging, thus making them unfit for direct 3D tomography. 3D geometric estimation from a large number of TIRF surface images captured at different incident angles has been recently reported \cite{yang2010estimation,boulanger2014fast}. These methods require solving inverse estimation problems and prior knowledge of the sample features. More recently, a simpler protocol of sequential imaging and photobleaching with multiangle TIRF to localize fluorescence emission to a region within the PSF of the objective was reported to achieve axial super-resolution \cite{fu2016axial}. 

\begin{figure}[t]
\centering
\includegraphics[width=1\linewidth]{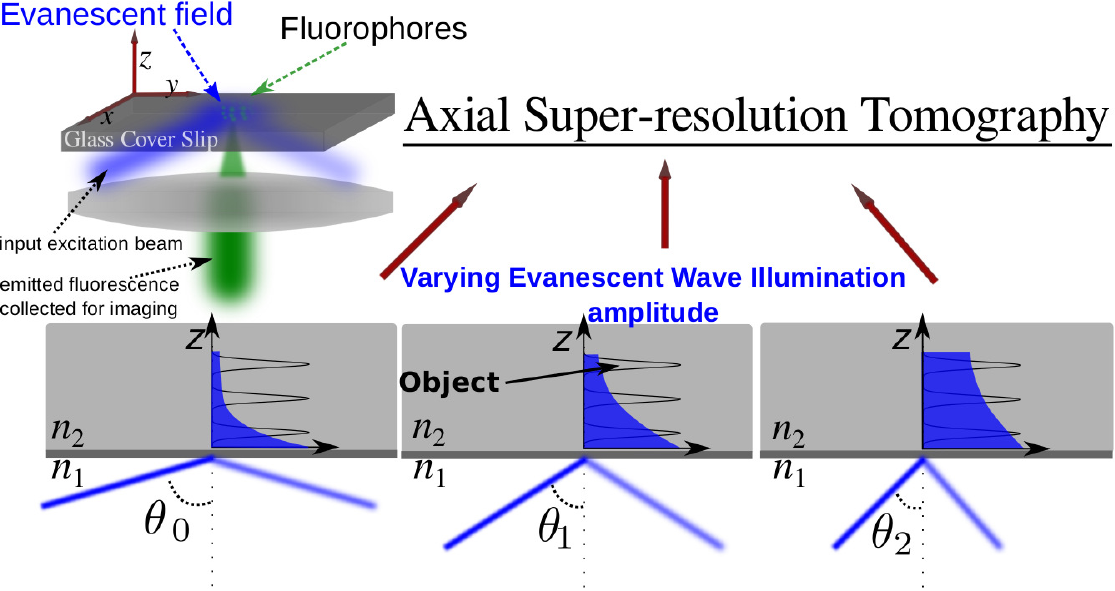}
\caption{Schematic of a TIRF microscope. Penetration depth of an evanescent wave can be controlled by the incident angle $\theta$ of the illuminating beam. As the incident angle is increased, the illumination depth decreases. When illumination depth of the sample increments in discrete steps, the information captured within the PSF of the objective lens also increments in steps, and the incremental information corresponds to super-resolution features.}
\label{fig:axial_superresolution_algo_concept}
\vspace{-10pt}
\end{figure}

In this paper we propose an axial super-resolution evanescent wave tomography (AxSET) method based on Total Internal reflection Fluorescence (TIRF) microscopy. We extract features with super-resolution from multiple sets of diffraction limited 3D images, where each set of 3D image is acquired at different illumination depths of the TIR evanescent wave. The proposed method does not require explicit knowledge of illumination depth or the PSF, nor does it rely on photobleaching of the sample. The method is tolerant to local variations in refractive index of the sample and angular bandwidth of the incident beam. We present the theoretical basis of our algorithm and show that it enables 3D tomography without additional optical components and sample preparation. We support our claims with 1D and 3D Fourier optics simulations and show that our algorithm can discern and reconstruct objects which otherwise appear coalesced with a conventional microscope. We experimentally validate our method by imaging and demonstrating super-resolution reconstruction of microtubules in HeLa cells. We demonstrate tomographic reconstruction with an axial resolution of $\sim$130nm (in contrast to $>$450~nm limit of the microscope). Although, the main goal of this paper is to increase the resolution in the axial direction, our method can be combined with other focal-plane super-resolution techniques like STORM and PSIM to achieve 3D super-resolution. 

\begin{figure}[t]
\centering
\includegraphics[width=1\linewidth]{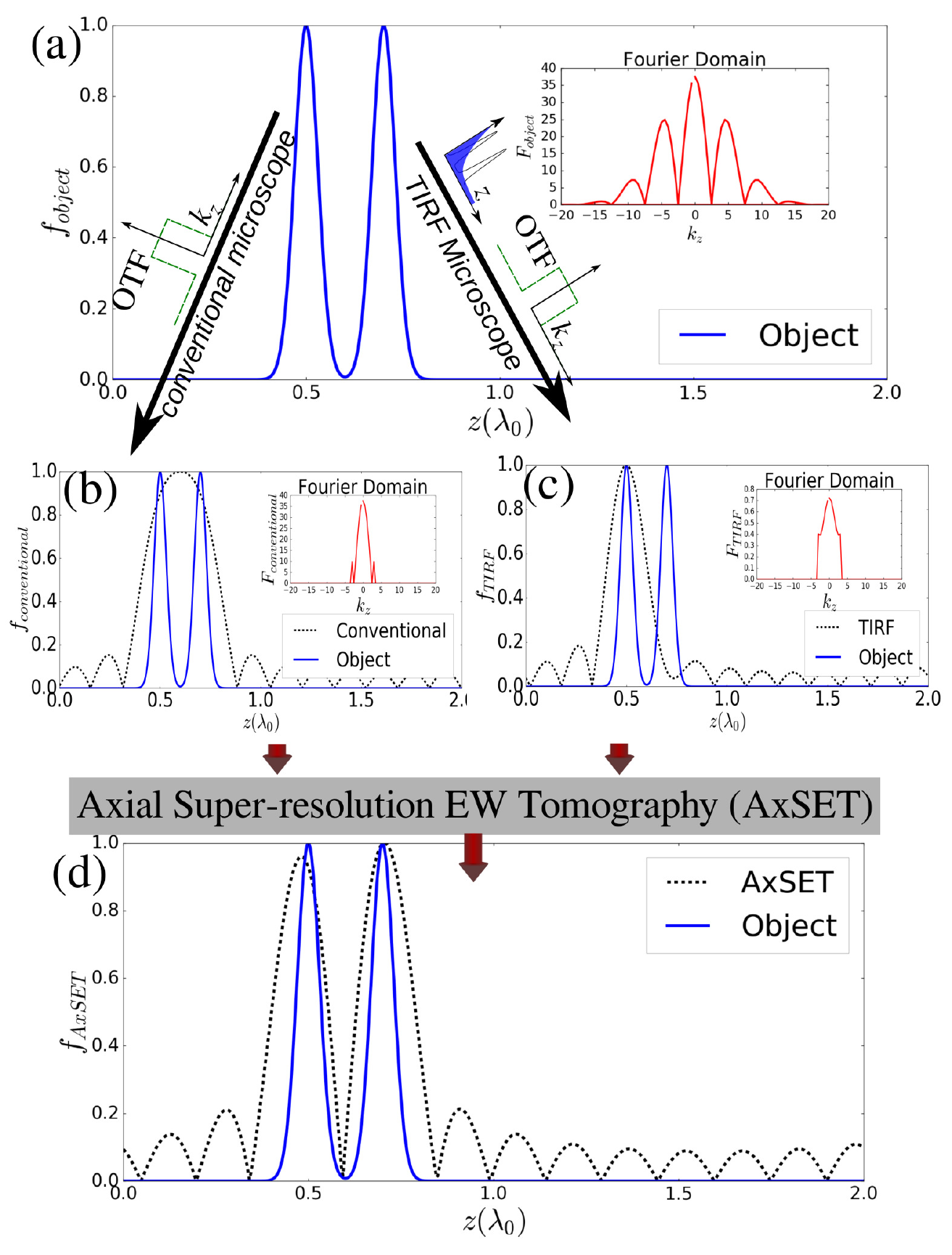}
\caption{Fourier reconstruction with confocal and TIRF image. $f_{obj}$ and $F_{obj}$ is the object in spatial domain and Fourier domain respectively. In conventional microscopy the object is passed through the Optical Transfer Function (OTF). The band-limited optical transfer function of the microscope introduces diffraction limit, and the resultant captured image $f_{conventional}$ is diffraction limited as shown in blue curves in left panel. The same object when imaged by TIRF is first illuminated with evanescent wave and the differentially illuminated object is then imaged by conventional system to get super-resolution $f_{TIRF}$. Fourier reconstruction is then performed on $F_{conventional}$ and $F_{TIRF}$ to extract and reconstruct the object with super-resolution.} 
\label{fig:block_diagram_Fourier_reconstruction}
\vspace{-10pt}
\end{figure}

The resolution limit arising from the ellipsoidal shape of the PSF of an optical microscope is $\Delta r \approx \lambda/(2 n \sin\phi)$ in the focal plane, and $\Delta z \approx \lambda/(n \sin^2\phi)$ along the optical axis \cite{hell2007far}. Here $\lambda$ is the wavelength of light in free space, $n$ is the refractive index of the medium, and $\phi$ is the aperture angle of the lens. The axial resolution, which is worse by a factor greater than two as compared to the focal-plane resolution, is significantly enhanced if instead of a propagating wave, the fluorescence in the sample is excited by an evanescent wave generated by total internal reflection at the interface of cover-slip and the sample, as shown in Fig.~\ref{fig:axial_superresolution_algo_concept}. The attenuation coefficient of an evanescent wave (EW) in the sample is given by $\alpha = (2\pi/\lambda)(n_1^2\sin^2\theta-n_2^2)^{(1/2)}$, where $n_1$ is the refractive index of the cover-slip and $n_2$ is the refractive index of the sample. $\theta$ is the angle between the incident light and normal to the interface. When the incident angle of light ($\theta$) is above the critical angle $\theta_c=\sin^{-1}(n_2/n_1)$, the wave is evanescent in the sample, with a theoretical upper limit to the attenuation being $\alpha_{max} = (2\pi/\lambda)(n_1^2-n_2^2)^{(1/2)}$. Since the attenuation $\alpha$ of the wave can be controlled between $0$ and $\alpha_{max}$ by the incident angle $\theta$, it is possible to allow illumination of a desired thickness and acquire a z-stack of the focal-plane images over the desired thickness, as depicted in Fig.~\ref{fig:axial_superresolution_algo_concept}. The 3D image thus obtained will have high SNR, but since the image is captured via a conventional optical microscope, the resolution of each focal plane image will be diffraction limited, and will be governed by an ellipsoidal PSF. Thus tomography with optical sectioning is diffraction limited in the axial direction even in EW illumination mode. However, due to evanescent wave illumination, near-surface features have higher amplitudes than those at deeper planes, and this reflects in the Fourier components of the corresponding features in the acquired images. By comparing the Fourier components of two images with different illumination depths, features at deeper planes can be identified at a resolution greater than the diffraction limit. In other words, axial super-resolution features can be extracted from the diffraction limited images by eliminating the near-surface super-resolution features captured in high-resolution TIRF illumination mode. We call this method axial super-resolution evanescent wave tomography (AxSET). 

The concept of AxSET is explained in Fig.~\ref{fig:block_diagram_Fourier_reconstruction}. Fig.~\ref{fig:block_diagram_Fourier_reconstruction}a shows the spatial distribution ($f_{object}(z)$) of the sample in the $z$ direction, and its Fourier representation ($F_{obj}(k_z)$) is shown in the inset. For illustration, we have considered a two particle object. Both the particles in the object as well as the separation between them leave their signature throughout the Fourier domain. Since the optical transfer function (OTF) of the microscope is band-limited, the Fourier representation of the image obtained by conventional imaging is truncated and the two particles appear coalesced in the spatial domain if they are closer than the resolution limit (black dashed curve in Fig.~\ref{fig:block_diagram_Fourier_reconstruction}(b)). This image is therefore diffraction limited. Fig.~\ref{fig:block_diagram_Fourier_reconstruction}(c) depicts the image acquired from the same object $f_{obj}$ in TIRF mode. Before the image is acquired by the optical microscope, the object is illuminated by an evanescent wave $f_{ev}=e^{-\alpha z}$, resulting in the differential amplitude of the particles at different depths, such that only the particle near the surface (z=0 plane) is prominent. Therefore the Fourier representation of the TIRF illuminated object has a signature of only the first particle. Note that the object and evanescent wave illumination multiply in the spatial domain and convolve in the Fourier domain. When this differentially  illuminated object is imaged by the optical microscope, the resultant truncation of the Fourier spectrum ($F_{TIRF}$) leads to elongation of the reconstructed particle ($f_{TIRF}$), but nevertheless, the particle can be distinctly identified at its correct position. This image, represented by $f_{TIRF}$ is therefore super-resolved near the surface. The difference between the conventional and TIRF image is that even in the band-limited Fourier representation, the conventional image has signatures from both the particles, while in TIRF the signature of the particle from the deeper layers is lost. We will call the signature which is lost in the TIRF but present in conventional fluorescence microscopy as $F_{residue}(k_z)$. The expression relating the TIRF and conventional images can then be written as:

\begin{eqnarray}
\label{eq:2} F_{conventional}(k_z) &=& C *  F_{TIRF}(k_z) + F_{residue}(k_z)
\end{eqnarray}
$C$  is a scaling factor, which in general can be a function of $k_z$. We solve eq(\ref{eq:2}) using a deconvolution operation between $F_{conventional}$ and $F_{TIRF}$ and compute the $F_{residue}$. $F_{residue}$ has super-resolution information, which when combined with $F_{TIRF}$, gives the AxSET reconstructed object as:

\vspace{-0.25cm}

\begin{equation}
F_{AxSET}(k_z) = F_{TIRF}(k_z) + F_{residue}(k_z)
\label{eq:3}
\end{equation}

The reconstructed object is represented by a black dashed line in Fig.~\ref{fig:block_diagram_Fourier_reconstruction}(d), and the original object is shown by the blue curve. It can be seen that while a conventional microscopy shows the two objects as coalesced and TIRF shows only the near-surface object, the AxSET reconstructed image distinctly shows both the objects. Note that the Fourier representation is in complex space where the phase of the signal has critical information. Eq(\ref{eq:3}) is therefore not equivalent to simple addition of the amplitudes. When there are multiple objects, we might need to extract the features sequentially for incrementally increasing penetration. The axial resolution limit of the proposed method is limited by the ability of the attenuating illumination to discern features between two adjacent layers of the sample. Therefore the achievable axial resolution will be higher near the surface and would approach the diffraction limit for deeper illumination. 

\begin{figure}[t]
\centering
\includegraphics[width=0.9\linewidth]{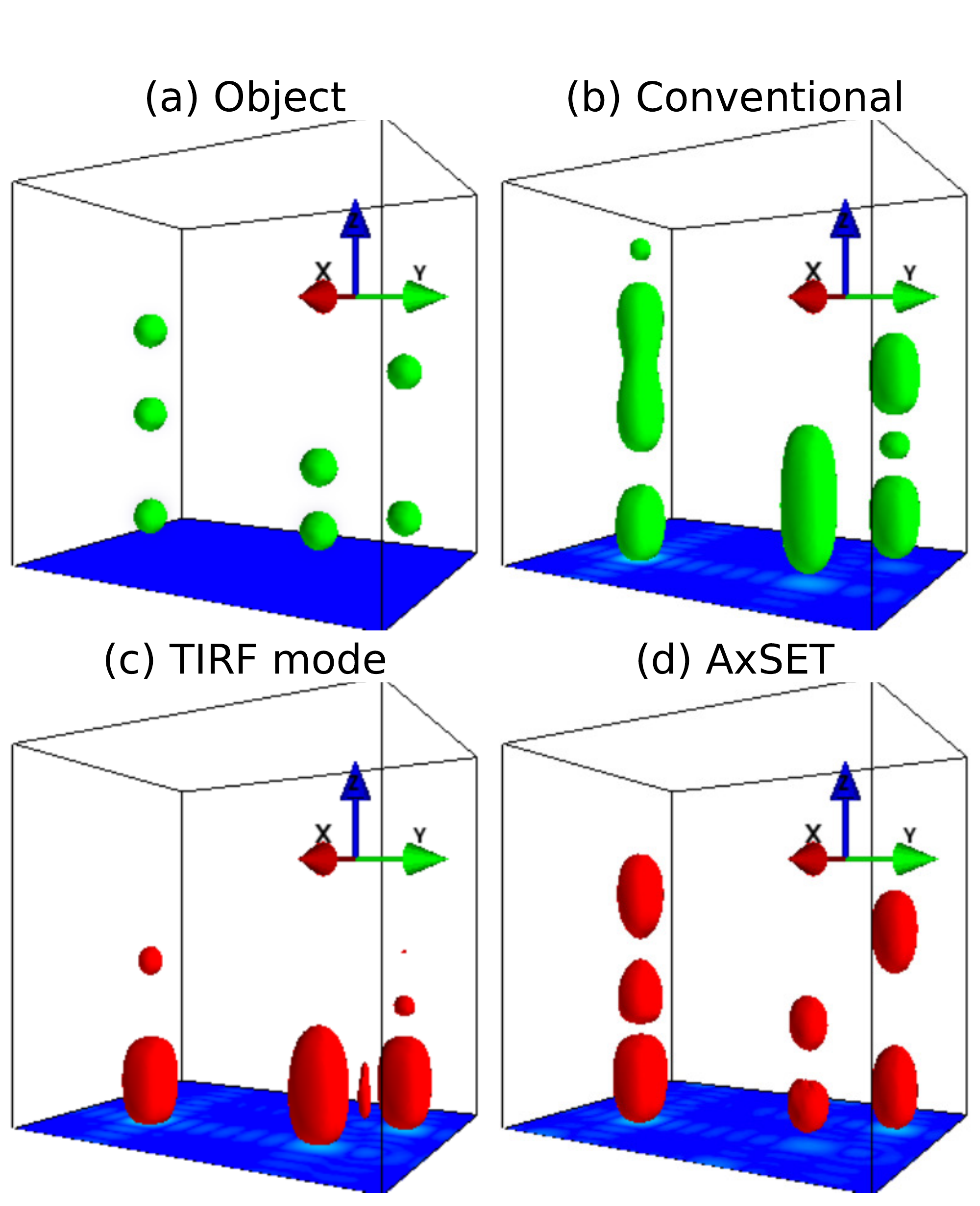}
\caption{AxSET reconstruction of a multi-particle 3D object. Panel (a) shows the digitally created object. (b) shows the 3D image of the object formed by the conventional imaging. (c) shows the 3D image formed in the TIRF surface mode, and (d) shows the super-resolution AxSET image formed by processing the (b) and (c) images.}
\label{fig:compare_3d_imaging}

\end{figure}

\begin{figure}[t]
	\centering
	\includegraphics[width=0.95\linewidth]{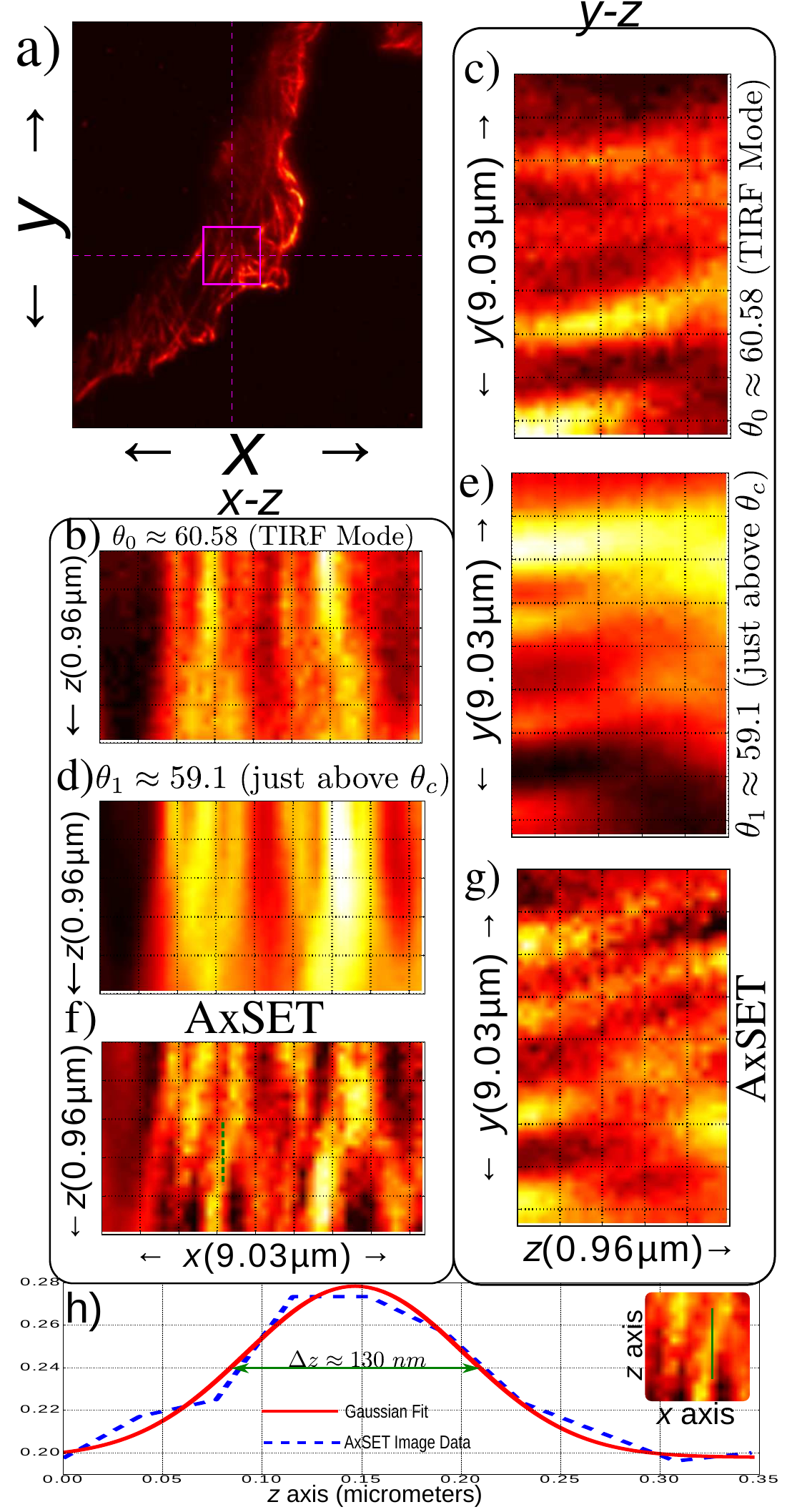}
	\caption{AxSET reconstruction of microtubule structures in HeLa cells stained with  alexafluor-488.  Panel (a) shows a $x-y$ plane optical section of the sample and the region of interest (ROI) is shown by solid magenta square. (b) and (c) show the TIRF mode $x-z$ and $y-z$ planes of the ROI at cross-wires(dashed magenta lines in (a) ), respectively. (d) and (e) show the corresponding $x-z$ and $y-z$ planes in conventional illumination mode. TIRF and conventional 3D images are processed by AxSET and the planes of reconstructed super-resolution image are shown in (f) and (g). (h) shows axial data of a mircotubule with a resolution of about 130~nm.}
	\label{fig:AxSET_set_experimental}
\end{figure}

Next, we implement AxSET to perform 3D tomography of a digitally created multi-particle object as shown in panel (a) of Fig.~\ref{fig:compare_3d_imaging}. The 3D image of the object from a conventional microscope when the full depth of the object is illuminated is shown in panel (b). It can be seen that in the 3D image, individual particles, which are all spheres, appear elongated and some particles are coalesced. The same object when imaged in the surface illumination mode gives the 3D reconstruction as shown in panel (c). In surface illumination mode, only the particles close to the surface (z=0 plane) are visible. For reconstruction, the actual distribution of the object is unknown. All we have is image (b) and (c). When the 3D images (b) and (c) are processed as per eq(\ref{eq:2},\ref{eq:3}), we get the 3D AxSET reconstruction as shown in Fig.~\ref{fig:compare_3d_imaging}(d). The reconstructed tomography matches well with the original distribution of particles in the object. It should be noted that the proposed AxSET method simultaneously generates high fidelity tomographic reconstruction of particles which are closer than the diffraction limit, as well as those that are scattered further away. Thus our method is suitable for super-resolution tomographic reconstruction of samples with a variety of features along the axial direction. It can be seen that the lower left particle in the AxSET reconstructed image is more elongated as compared to the other reconstructed particles. This is because, along the line of three particles, the AxSET reconstruction with two the image-sets extracts super-resolution features of only the deeper plane particles. Better resolution can be achieved with more sets of 3D images.

We validate our method on microtubule structures in HeLa cells tagged with  alexafluor 488 fixed on a glass coverslip ($n_1=1.55$). The cells are immersed in an aqueous medium ($n_2=1.33$). Fig.~\ref{fig:AxSET_set_experimental}(a) shows the $x-y$ plane optical section of the sample. The imaging was performed on Nikon eclipse Ti TIRF microscope with 100X objective with 1.49 numerical aperture. The sample is imaged at two illumination angles, $\theta_0\approx60.58^\circ$ for surface illumination, and $\theta_1\approx 59.1^\circ$ (close to critical angle) for deeper illumination. The $x-z$ and $y-z$ plane axial sections of the 3D image at $\theta_0$ illumination is shown in Fig.~\ref{fig:AxSET_set_experimental}(b)-(c) respectively, while Fig.~\ref{fig:AxSET_set_experimental}(d)-(e) shows the corresponding $x-z$ and $y-z$ plane axial sections for $\theta_1$ illumination, respectively. Processing these 3D images with the proposed AxSET method yields a super-resolution image whose axial sections are shown in Fig.~\ref{fig:AxSET_set_experimental}(f)-(g). It can be seen that the features which were blurred and smeared out in the raw images can be seen with a precision of around 130~nm in the reconstructed image. Fig.~\ref{fig:AxSET_set_experimental}h shows the cut-through of AxSET image along one microtubule.The resolution of the microscope is 210~nm in the focal plane and $\sim$450~nm along the optical axis. The processed image clearly shows resolution enhancement by a factor of $\sim$3 along the $z$-axis.  

The proposed AxSET enables full 3D tomographic reconstruction with axial resolution higher than the optical acquisition system allows. This method can be employed to extract super-resolution features wherever an evanescent wave can be used to excite the sample. This method can be combined with focal plane super-resolution techniques like STORM and will find applications in wide range of 3D microscopy techniques ranging from optical to THz imaging.


The authors are grateful to the Institute of Oil-Sands Innovation (IOSI), Canada, for the funding. The authors thank Gareth Lambkin from the Biological Services, Department of Chemistry, University of Alberta, for technical support.

\bibliographystyle{ieeetr}

\bibliography{references_EWT}

\appendix


\section{1D  simulation of AxSET}
To demonstrate super-resolution in the axial direction, we simulate the optical imaging of a one dimensional object $f_{obj}(z)$ (distribution along axial direction) for a conventional imaging, TIRF imaging and the AxSET reconstruction. We define particles in the object by a Gaussian distribution,

\begin{equation}
f_{obj} = \sum_{i=1}^N e^{-\frac{(z-z_i)^2}{2\sigma^2}}
\label{eq:supp1}
\end{equation}
where $z_i$ is the position of the $i^{th}$ object, and $\sigma$ defines its width. For the simulation of conventional imaging, we first take the Fast Fourier transform (FFT) of $f_{obj}(z)$ using the inbuilt function in numpy \cite{van2011numpy} package in Python. The resultant FFT of the object ($F_{obj}(k_z)$) is then multiplied by the optical transfer function (OTF) of the form

\begin{equation}
OTF(k_z) = \left\{ \begin{array}{l c}
1 ; & -k_c< k_z < k_c  \\
0 ; &\mathrm{else}
\end{array} \right.
\label{eq:supp2}
\end{equation}
Here $k_c$ is the cut-off frequency of the optical transfer function. If $\Delta z$ is the axial resolution limit, then $k_c \approx 1/\Delta z$. The Fourier domain image from the conventional microscope is then obtained by,
\begin{equation}
F_{conventional}(k_z) = OTF(k_z)\cdot F_{obj}(k_z)
\label{eq:supp3}
\end{equation}
Image of the object ($f_{conventional}(z)$) in the spatial domain is recovered by taking the inverse FFT of $F_{conventional}(k_z)$.

In the TIRF microscopy simulation, the object is first multiplied by an evanescent wave (EW) illumination represented by $f_{ev}(z) = e^{-\alpha z}$ ($\alpha$ being the attenuation coefficient of EW), before conversion to the Fourier domain. The EW-illuminated object is then multiplied by the OTF of a microscope to get the Fourier domain representation of the TIRF image. 
\begin{equation}
F_{TIRF}(k_z) = OTF(k_z)\cdot(F_{obj}(k_z)*F_{ev}(k_z))
\label{eq:supp4}
\end{equation}
The inverse FFT of $F_{TIRF}(k_z)$ gives the spatial domain TIRF image.

As explained in the main text of the paper, we represent the signature of a particle lost in the TIRF but present in the conventional image as $f_{residue}(z)$. The expression relating the TIRF and the conventional images can then be written as:

\begin{eqnarray}
\label{eq:supp5} f_{conventional}(z) &=& c\cdot f_{TIRF}(z) + f_{residue}(z)\\
\label{eq:supp6} F_{conventional}(k_z) &=& C *  F_{TIRF}(k_z) + F_{residue}(k_z)
\end{eqnarray}
where $c$ is a scaling factor, which in general can be a function of $z$. However, here we consider it to be a scalar (variation in the $OTF$ of the objective over $z$ is negligible). We solve equation~(\ref{eq:supp6}) with a  deconvolution function implemented in Scientific Python computation package scipy \cite{scipy}, to get $F_{residue}({k_z})$. The object is then reconstructed to get AxSET in the Fourier domain,

\begin{equation}
F_{AxSET}(k_z) = F_{TIRF}(k_z) + F_{residue}(k_z)
\label{eq:supp7}
\end{equation}
AxSET image in the spatial domain is then recovered by taking the inverse FFT of $F_{AxSET}(k_z)$.

The parameters of simulation for Fig.~1 in the main text of the paper are: $z_1=0.5$, $z_2=0.7$, $\sigma=0.03$, $k_c=3$, and $\alpha=7$. 

\section{Visualization of 3D AxSET}

For simulation of the 3D imaging, we follow the same procedure as described above, but the spatial and Fourier domain is now three dimensional. The 3D object is defined by,

\begin{equation}
f_{obj} = \sum_{i=1}^2 e^{-\frac{(x-x_i)^2+(y-y_i)^2+(z-z_i)^2}{2\sigma^2}}
\label{eq:supp8}
\end{equation}
and the OTF is defined as,

\begin{equation}
OTF(k_x,k_y,k_z) = \left\{ \begin{array}{l c}
& (-k_{cx}< k_x < k_{cx}) \\ 
1 ; & (-k_{cy}< k_y < k_{cy}) \\ &  (-k_{cz}< k_z < k_{cz})  \\
0 ; &\mathrm{else}
\end{array} \right.
\label{eq:supp9}
\end{equation}
The $k_{cx}$, $k_{cy}$ and $k_{cz}$ are cut-off spatial frequency of OTF, such that $k_{cx}=k_{cy}=2k_{cz}$. The subsequent procedure for AxSET reconstruction remains the same. 

Visualization of the 3D object and the corresponding 3D images is performed in Mayavi visualization package \cite{ramachandran2011mayavi} in Python, using the isosurface module. 

\section{Biological sample preparation and imaging}
HeLa Cells were grown on coverslips (Number 1.5, Electron Microscopy Sciences, Hatfield, PA) in DMEM plus 10\% fetal calf serum. Cells were fixed in PBS containing 3.7\% paraformaldehyde and 0.2\% gluteraldehye for 12 minutes. Following reduction with 0.2\% sodium borohydride, cell membranes were permeabilized with 0.2\% Triton X-100 and non-specific binding was blocked using 3\% BSA in PBS for 30 minutes. Coverslips were subsequently incubated with primary anti-beta tubulin (T5201, Sigma-Aldrich) diluted 1:500 in PBS containing 3\% BSA and 0.2\% Triton X-100 for one hour. After washing with PBS containing 0.2\%BSA and 0.1\% Triton X-100, the coverslips were incubated with a donkey anti-mouse secondary antibody conjugated to alexafluor 488 (Jackson Immuno Research Labs, West Grove, PA) for 30 minutes. After removing excess secondary by 3 rounds of washing in PBS, antibodies were post-fixed by incubation with 4\% paraformaldehyde in PBS for 10 minutes followed by washing with PBS.

The imaging was performed on HeLa cells in aqueous medium using a Nikon eclipse Ti TIRF microscope with a 100X objective with numerical aperture 1.49. A 488~nm laser (Melles Griot) was used for excitation of the sample, and the emitted fluorescence was collected via a 500-550~nm filter set. The Fluorescence images were captured by a QuantEM CCD camera.

3D images were formed by acquiring a z-stack of the $x-y$ plane optical sections, with an axial distance between each section being $0.04~\mu m$. Two sets of 3D images were captured, one with deep illumination acquired at an incident angle closer to the critical angle, and other in the TIRF mode. The image processing is handled with scikit-image \cite{van2014scikit} library in Python.

\section{More experimental images from AxSET}

Here we present few additional experimental images to show the AxSET reconstruction. This is to reinforce our claim that the AxSET is suitable for the reconstruction of a variety of features.

\begin{figure}[h]
	\centering
	\includegraphics[width=0.95\linewidth]{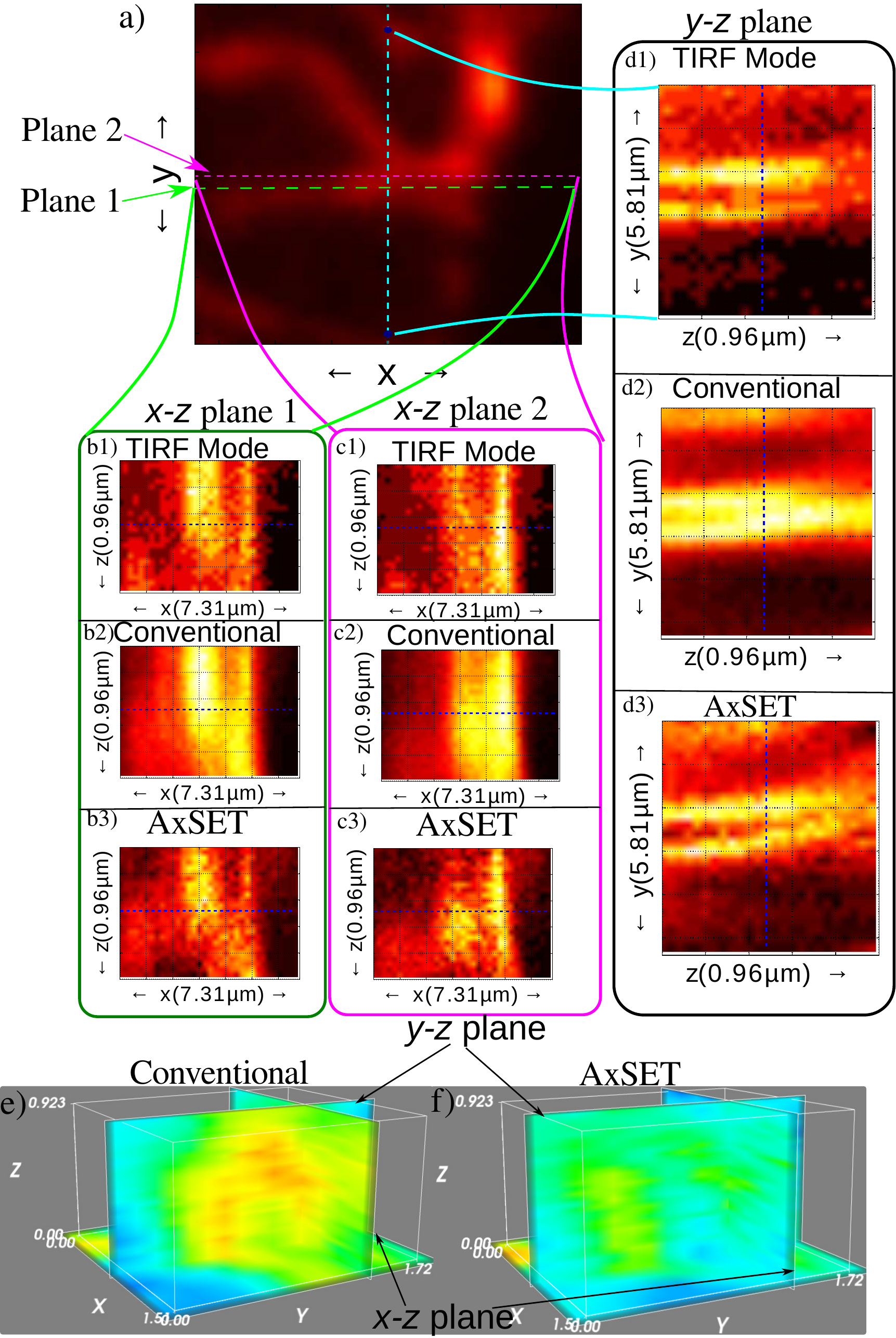}
	\caption{AxSET reconstruction of microtubules in a HeLa cell. Panel (a) shows the $x-y$ plane optical section of a region in the sample. One $y-z$ plane is marked by dashed cyan line while the two $x-z$ planes are marked by green (plane 1) and magenta (plane 2) dashed lines respectively. Panel (b1,b2,b3) show the $x-z$ plane~1 in TIRF, conventional and AxSET reconstructed 3D images, respectively. Similarly panel (c1,c2,c3) show the TIRF, conventional and AxSET reconstructed 3D images at $x-z$ plane~2. Panel (d1,d2,d3) depicts the TIRF, conventional and AxSET images of $y-z$ plane, respectively. It can be observed in both the $x-z$ and $y-z$ planes that the microtubules which appeared coalesced in conventional imaging (and were lost at deeper planes in TIRF) can be reconstructed distinctly by AxSET. In panel (b) and (c), the super-resolved microtubules can be seen moving with between the two $x-z$ planes as the $y$ coordinates of the axial optical section change. Plane~1 and Plane~2 are separated by a distance of 210~nm. Here, an AxSET resolution of about 130~nm can be observed. Panel (e) and (f) compares the 3D image plane cross-sections in conventional and AxSET reconstruction.}
	\label{fig:AxSET_set_11-5-6_with_3D_imageplanes}
\end{figure}

Fig.~\ref{fig:AxSET_set_11-5-6_with_3D_imageplanes} shows that AxSET can capture the change in position of microtubules in axial direction between two adjacent axial planes. 

\begin{figure}[h]
	\centering
	\includegraphics[width=1\linewidth]{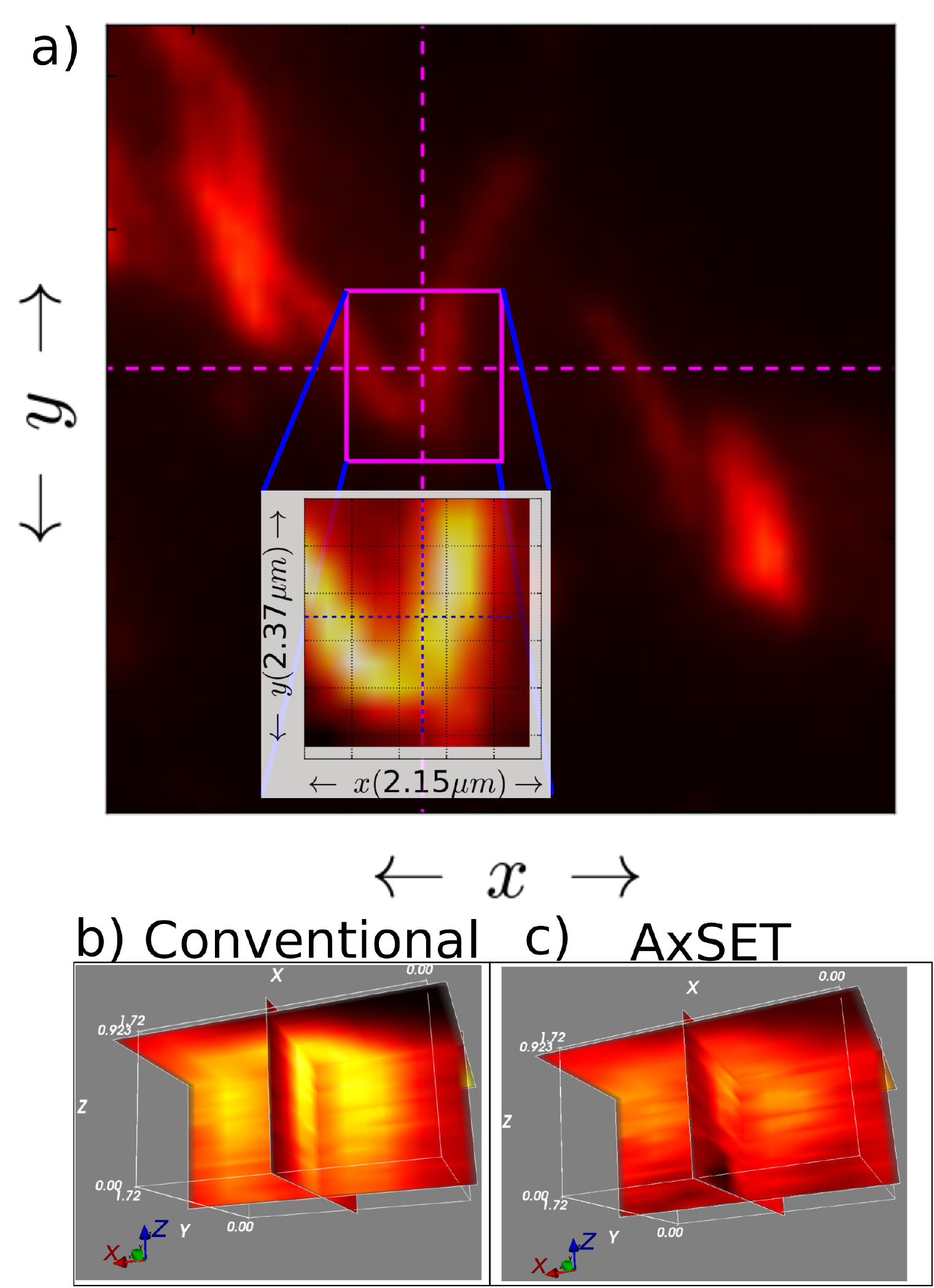}
	\caption{3D image planes of the AxSET reconstruction of a curving microtubule. (a) is the $x-y$ plane optical section of the microtubule. A smaller region of interest over which AxSET is performed is shown in the inset. Panel (b) and (c) show the three image plane in the conventional and AxSET  3D images. Thickness of the microtubules in the AxSET image planes show significant improvement in the resolution over conventional image.}
	\label{fig:AxSET_set11-9_with_mayavi_3d}
\end{figure}

Fig.~\ref{fig:AxSET_set11-9_with_mayavi_3d} shows the AxSET reconstruction of a curving microtubule in a HeLa cell with the help of 3D image planes. The orthogonal image planes shows that the axial location of the microtubule is better resolved in AxSET. 

\begin{figure}
	\centering
	\includegraphics[width=1\linewidth]{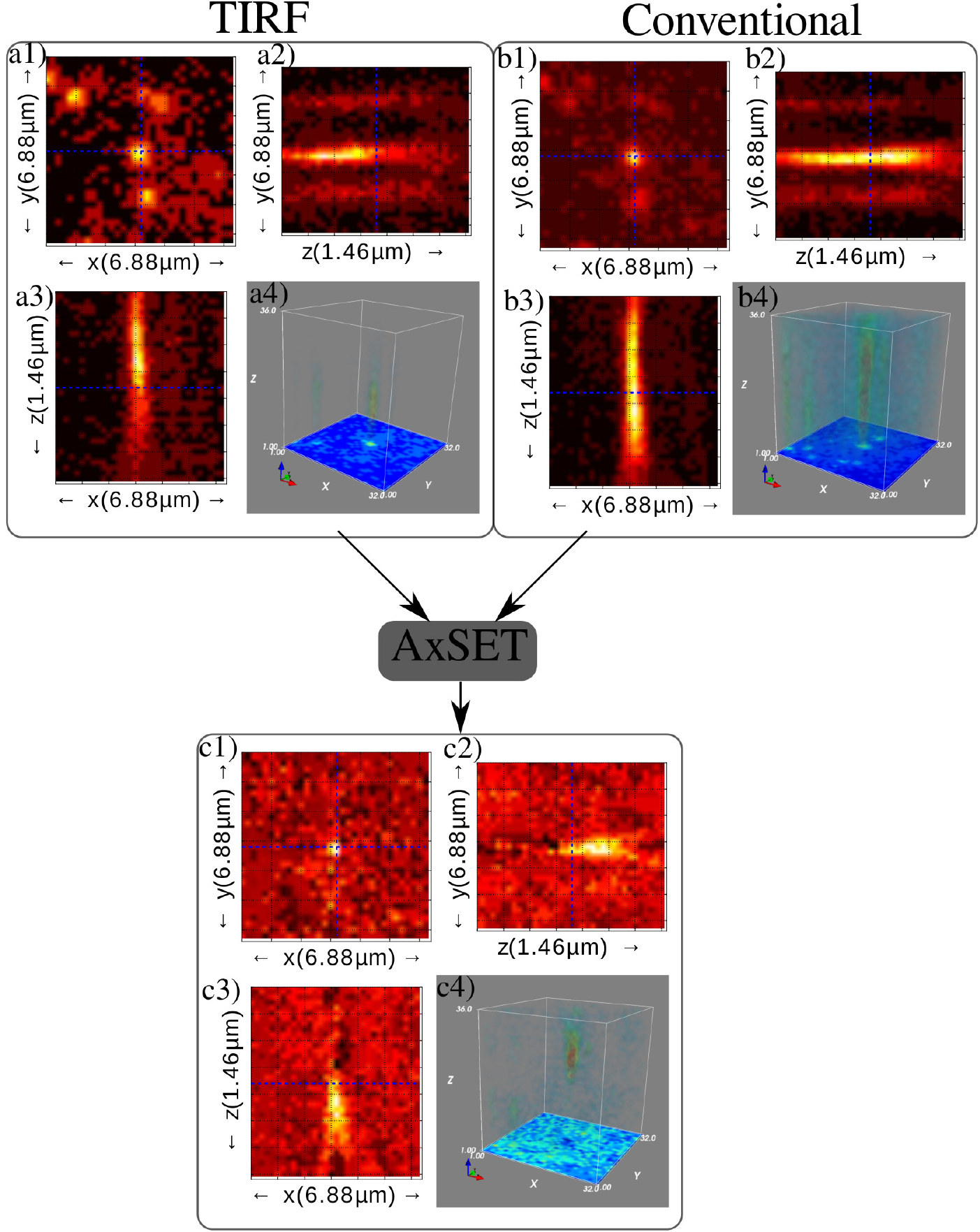}
	\caption{AxSET reconstruction of fluorescent particle from TIRF and conventional 3D images. Panel (a) shows the optical cross-sections of 3D image acquired in TIRF mode. (a1) is the $x-y$ plane cross-section, (a2) is the $y-z$ plane cross-section at the cross-wires (blue dashed lines in $x-y$ plane) and (a3) is the $x-z$ plane. Panel (a4) is the 3D visualization of the image. Similarly, panel (b) represent the corresponding 3D image captured in conventional mode (deep illumination). 3D images (a) and (b) are processed by AxSET method to get the super-resolution reconstructed image, is shown in panel (c). It can be seen that AxSET reconstruction has much better resolution ($\sim150$~nm) in the $z$-direction. The inset 3D visualization is in the units of pixels, and not physical coordinates. Physical dimensions are computed from the pixels by multiplying the pixel index by 210~nm in the $x-y$ plane and by 40~nm in the $z$-direction.}
	\label{fig:AxSET_set21-3}
	\vspace{-10pt}
\end{figure}

Fig.~\ref{fig:AxSET_set21-3} shows the AxSET reconstruction of a fluorophore particle, which is very similar to a point particle. Comparison of AxSET with conventional 3D image conclusively show that AxSET reconstruction is about 3 times better resolved.



\end{document}